\documentclass[floats,floatfix,showpacs,amssymb,prd,twocolumn,nofootinbib,longbibliography]{revtex4-1}

\usepackage{anyfontsize}
\usepackage{amssymb}
\usepackage{amsmath}
\usepackage{amsfonts}
\usepackage{amsbsy}
\usepackage[utf8]{inputenc} 
\usepackage{newtxtext}
\usepackage{newtxmath}
\usepackage{enumerate}

\usepackage{tensor}
\usepackage{mathrsfs}

\usepackage{bm}

\usepackage[utf8]{inputenc}

\usepackage{graphicx}
\usepackage{epsfig}
\usepackage{epstopdf}

\usepackage[usenames,dvipsnames]{xcolor}

\usepackage{verbatim,mathtools,needspace,enumitem,etoolbox}

\definecolor{linkcolor}{rgb}{0.0,0.3,0.5}

\usepackage[unicode, colorlinks=true, linkcolor=linkcolor, citecolor=linkcolor, filecolor=linkcolor,urlcolor=linkcolor, pdfusetitle]{hyperref}

\usepackage{epstopdf}

\usepackage{bm}
\usepackage{dcolumn}

\usepackage{longtable}
\usepackage{enumerate}
 \usepackage[normalem]{ulem}
\definecolor{darkgreen}{RGB}{34,139,34}

\definecolor{romared}{RGB}{142,0,28}
\hypersetup{colorlinks=true,
            citecolor=romared,
            linkcolor=romared,
            urlcolor=romared}
\usepackage[T1]{fontenc}
\usepackage{subfigure}
\usepackage{amsmath}
\usepackage{float}
\usepackage{mathrsfs}
\usepackage[dvipsnames]{xcolor}
\usepackage{soul}

\setcitestyle{numbers,square}

\begin{document}
\title{Spinning black holes with a separable Hamilton-Jacobi equation \\ from a modified Newman-Janis algorithm
}

\author{Haroldo C. D. Lima Junior}
\email{haroldo.ufpa@gmail.com} 
\affiliation{Faculdade de F\'{\i}sica,
Universidade Federal do Par\'a, 66075-110, Bel\'em, PA, Brasil }

\author{Lu\'{\i}s C. B. Crispino}
\email{crispino@ufpa.br} 
\affiliation{Faculdade de F\'{\i}sica,
Universidade Federal do Par\'a, 66075-110, Bel\'em, PA, Brasil }

\author{Pedro V. P. Cunha}%
 \email{pedro.cunha@aei.mpg.de}
\affiliation{%
 Max Planck for Gravitational Physics - Albert Einstein Institute, Am M\"{u}hlenberg 1, Potsdam 14476, Germany
}%

\author{Carlos A. R. Herdeiro}
\email{herdeiro@ua.pt} 
\affiliation{Departamento de Matem\'atica da Universidade de Aveiro and Centre for Research and Development  in Mathematics and Applications (CIDMA), Campus de Santiago, 3810-183 Aveiro, Portugal }

\begin{abstract}
Obtaining solutions of the Einstein field equations describing spinning compact bodies is typically challenging.
The Newman-Janis algorithm provides a procedure to obtain rotating spacetimes from a static, spherically symmetric, seed metric. It is not guaranteed, however, that the resulting rotating spacetime solves the same field equations as the seed. Moreover, the former may not be circular, and thus expressible in Boyer-Lindquist-like coordinates. 
Amongst the variations of the original procedure, a modified Newman-Janis algorithm (MNJA) has been proposed that, by construction, originates a circular, spinning spacetime, expressible in Boyer-Lindquist-like coordinates. As a down side, the procedure introduces an ambiguity, that requires extra assumptions on the matter content of the model.  In this paper we observe that the rotating spacetimes obtained through the MNJA \textit{always} admit separability of the Hamilton-Jacobi equation for the case of null geodesics, in which case, moreover, the aforementioned ambiguity has no impact, since it amounts to an overall metric conformal factor. {We also show that the Hamilton-Jacobi equation for light rays propagating in a plasma admits separability if the plasma frequency obeys a certain constraint.} As an illustration, we compute the shadow and lensing of some spinning black holes obtained by the MNJA.
\end{abstract}

\maketitle

\section{Introduction}
\label{SecI}
The Event Horizon Telescope has recently released the first observed black hole shadow~\cite{EHT}. This observation concerns the supermassive black hole located at the center of the M87 galaxy. Observational evidence suggests that this black hole, as well as many others, has non-vanishing angular momentum. Historically, the first rotating black hole of General Relativity, as well as the most influential one, is the Kerr solution~\cite{Kerr:1963}. Remarkably, the Kerr solution, which is the most general physically reasonable black hole of vacuum General Relativity,  is fully described by solely two global charges~\cite{Carter:1971zc}: the black hole mass and angular momentum.

In general, exact rotating black hole solutions are difficult to obtain by simply inserting a sufficiently general ansatz into the Einstein field equations. The resulting set of non-linear, coupled, partial differential equations is prohibitively difficult, even though it can often be solved numerically. The Kerr solution was obtained by further assuming that the metric should be algebraically special, which, via the Goldberg-Sachs theorem~\cite{GS}, implied the existence of special congruence of curves  and simplified the equations. Shortly afterwards, it was observed by Newman and Janis that a certain \textit{ad hoc} algorithm yielded a simple method to obtain the Kerr solution, a procedure that became known as the  Newman-Janis algorithm (NJA)~\cite{NJA:Kerr}. The procedure is based on a certain complexification, a solution generating technique used in other contexts since the XIXth century - see $e.g.$~\cite{Ciotti:2008tn} and references therein. The NJA starts with the static, spherically symmetric Schwarzschild geometry; after implementing the complexification, which in particular introduces an extra parameter, the Kerr solution emerges as a result. A success of the NJA was the original derivation of the Kerr-Newman solution~\cite{NJA:Kerr-Newman}, obtained by applying the complexification procedure to the Reissner-Nordstr\"om electrovacuum black hole.

Albeit a useful trick, two (related) caveats overshadow the NJA. Firstly, why does it work? Only some partial insights have been offered, see $e.g.$~\cite{NJA:Kerr,Newman:1973yu,Drake:1998gf}. Secondly, when does it work? Although there are rotating solutions obtained via the NJA from static solutions that can fall within the same model, this is not always the case. Sometimes, the resulting rotating metric fails to obey the field equations of the seed model. An example where the NJA fails is the four dimensional Einstein-Maxwell-dilaton black hole, whose spinning generalization cannot be obtained via the NJA~\cite{Yazadjiev:1999ce} (and in fact it is unknown, in closed form, for generic dilaton coupling). Another example of the limited scope of validity of the NJA occurs in higher dimensions. The NJA can be used to transform the Tangherlini black hole~\cite{Tangherlini:1963bw} into a Myers-Perry black hole~\cite{Myers:1986un}, see~\cite{Xu:1988ju,Erbin:2014lwa}; but it fails to generate the spinning black holes of the higher dimensional Einstein-Maxwell model (without Chern-Simons terms), which again are still not known in closed analytic form. 

Despite these two issues, the NJA has been adopted as a useful technique in a relativist's toolkit to generate spinning metrics. These issues, moreover, are mitigated in cases where the static seed metric is, itself, postulated \textit{ad hoc}, as a case study to investigate some geometrical possibility, $e.g.$ a regular black hole. After generating the spinning geometry, one may then look for some matter model such that the spacetime solves the corresponding Einstein equations. 

Even taking this pragmatical perspective, however, there is yet another property of the NJA that is unsatisfactory, at least for some applications. For some seed metrics, the NJA results in a rotating spacetime that is not circular. In particular, this means it cannot be written in the canonical coordinates~\cite{NJA_mod0}. For Kerr these are the standard Boyer-Lindquist coordinates~\cite{Boyer:1966qh}. We shall call the corresponding coordinates \textit{Boyer-Lindquist-like} coordinates. Specifically, these are coordinates adapted to the Killing symmetries with only one off diagonal term, occurring in the Killing sector. This inability turns out to be related to the specific implementation of the complexification procedure. 

A way around the latter issue was recently proposed in the form of 
a \textit{modified} NJA (MNJA)~\cite{NJA_mod,NJA_mod2}. This procedure modifies the original NJA in such a way that it guarantees that a Boyer-Lindquist-like form emerges at the end. The price to pay is the emergence of an extra metric function, essentially unconstrained by the procedure, and therefore an ambiguity. In the MNJA it is proposed that this function should be fixed via physical arguments, such as the use of a particular form for the stress-energy tensor~\cite{NJA_mod}. As we shall see, however, for the purpose of understanding the null geodesic flow in the spinning geometries obtained from the MJNA, this ambiguity is irrelevant.  For examples of rotating solutions that have been obtained using the MNJA, see $e.g.$~\cite{MNJ_Sol_I,MNJ_Sol_II,MNJ_Sol_III,MNJ_Sol_IV,MNJ_Sol_V,MNJ_Sol_VI,MNJ_Sol_VII,MNJ_Sol_VIII}.

The gravitational field in the vicinity of a black hole is so strong that light rays can become trapped in bound orbits around the black hole. In general these are called fundamental photon orbits~\cite{Cunha:2017eoe}. Such orbits, of which light rings are a particular example, are closely related to the shadow cast by a black hole~\cite{Cunha:2018acu}. The study of the shadow of a Schwarzschild black hole was pioneered in Ref.~\cite{Sch_Shadow}, albeit the terminology shadow was only much later introduced~\cite{Falcke:1999pj}, while the shadow for a Kerr black hole was first correctly analyzed in Ref.~\cite{Kerr_shadow}. 
More recently, the shadows of many others rotating black hole solutions have been investigated, including, $e.g.$ of Kerr-Newman  black holes~\cite{Kerr_Newman_Shadow,Rotating_regularBH_Shadow}, Kerr black holes with scalar hair~\cite{KerrwSH_Shadow,Cunha:2016bjh,Cunha:2019ikd}, Kerr-Sen~\cite{Kerr_Sen_Shadow}, spontaneously scalarised Kerr black holes~\cite{Cunha:2019dwb}, Kaluza-Klein dyonic rotating black holes~\cite{Cunha:2018uzc}, the double Kerr solution~\cite{Cunha:2018cof}, spinning Einstein-Maxwell dilaton black holes~\cite{Cunha:2016wzk} and rotating regular black holes~\cite{Rotating_regularBH_Shadow}.

Studying the shadow of a given black hole, as well as the null geodesic flow in general, is considerably facilitated by the separability of the corresponding Hamilton-Jacobi equation. In the Kerr case, the separability of the Hamilton-Jacobi equation is a remarkable property first pointed out by Carter \cite{Carter::1968}. In this case, the geodesic equations of motion can be written as a set of 4 (first) order differential equations. The separability of Hamilton-Jacobi equation is related to the existence of a non-trivial (irreducible, rank 2) Killing tensor for the Kerr geometry, which gives rise to the well-known Carter constant~\cite{Carter::1968}. {Surprisingly, the Hamilton-Jacobi equation for light rays propagating in a plasma in Kerr spacetime also admits separability, if the frequency of the plasma satisfies a given constraint~\cite{Kerr_Plasma}. It is important to state that light rays in plasma do not follow geodesic curves, therefore the existence of a Carter-like constant is unexpected.} Besides the Kerr solution, many others rotating black hole solutions also admit separability of the Hamilton-Jacobi equation {for null geodesics} \cite{Kerr_Newman_Shadow,Kerr_Sen_Shadow,Rotating_regularBH_Shadow}.  Recently, it was shown that a rotating solution obtained from the NJA only admits separability for null geodesics, if the resulting metric can be written in
Boyer-Lindquist-like coordinates~\cite{NJA_shadow}. 
 It was also pointed out that the Hamilton-Jacobi equation in a spacetime obtained with the NJA admits separability of timelike geodesics if one of the metric function is additively separable~\cite{Che-Yu Chen:2019}. Moreover deformed rotating BH spacetimes have also been shown to admit separability of Hamilton-Jacobi equation~\cite{Deformed_BH_Separability}. Then, one may ask if a generic rotating configuration obtained through the MNJA also admits separability of the Hamilton-Jacobi equation and what is the role played by the unconstrained function introduced in the MNJA on the separability of null geodesics and light rays propagating in plasma. 

The aim of this paper is to stress that null geodesics in a spacetime generated through the MNJA \textit{always} admit separability, regardless of the ambiguity raised by the undetermined function that the MNJA introduces. This goes beyond the results obtained in Ref.~\cite{NJA_mod} in three ways. Firstly, a specific form of the metric was chosen in Ref.~\cite{NJA_mod} whereas we work with the most generic form generated by the MNJA. Secondly, we show this separability is insensitive to the ambiguity introduced in the MNJA,  which is solved in Ref.~\cite{NJA_mod} by advocating a specific form of the energy-momentum tensor; our results show that regardless of this choice, separability holds. Thirdly, this universality of the separability only holds for null geodesic, not timelike ones, a distinction not made in~Ref.~\cite{NJA_mod}. We also show that the Hamilton-Jacobi equation for light rays propagating in a cold, pressureless and non-magnetized plasma admits separability, if the plasma frequency satisfies a given constraint, which depends on the ambiguity introduced by the modified NJ algorithm. In addition, for illustrative purposes, we analyze the shadow cast by some rotating spacetimes obtained through the MNJA.

{We would like to further emphasize that the construction of new rotating geometries via a Newman-Janis algorithm is a very active field of research in the literature, partially motivated by the difficulty of finding exact rotating analytical solutions to generic matter models. As such, general results on the properties of null geodesics and their integrability on geometries generated by Newman-Janis algorithms are both timely and of interest to a wide research community.}

In Sec.~\ref{SecII}, we reproduce the MNJA steps in order to obtain a rotating spacetime geometry from a given spherically symmetric spacetime. In Sec.~\ref{SecIII}, we analyze the Hamilton-Jacobi equation for null geodesics \textit{in the most generic spacetime} obtained through the MNJA and show that the resulting equations are always separable. {In Sec.~\ref{SecPlasma}, we analyze the Hamilton-Jacobi equation for light rays propagating in a plasma and show that the resulting equations admit separability if the plasma frequency satisfies a certain constraint.} In Sec.~\ref{SecIV}, we perform a generic analysis of the shadow edge in such rotating spacetimes for null geodesics, and then present some illustrative concrete cases. We present our final remarks in Sec.~\ref{section:Conclusions}. We use natural units, $G=c=1$, and  metric signature (-\ ,+\ ,+\ ,+).

\section{Modified Newman-Janis algorithm}
\label{SecII}
Let us start by reviewing the MNJA proposed in~\cite{NJA_mod,NJA_mod2}. We start with a spherically symmetric seed metric, written as
\begin{equation}
\label{linel}ds^2=-G(r)dt^2+\frac{1}{F(r)}\,dr^2+H(r)(d\theta^2+\sin^2\theta\,d\phi^2) \ .
\end{equation}
This choice does not exhaust the gauge freedom. Next, we write the line element \eqref{linel} in advanced null coordinates, using
\begin{equation}
dt=\frac{1}{\sqrt{F\,G}}\,dr+du,
\end{equation}
resulting in 
\begin{equation}
\label{EDF_coord}ds^2=-G\,du^2-2\,\sqrt{\frac{G}{F}}\,dr\,du+H\,(d\theta^2+\sin^2\theta\,d\phi^2)\ .
\end{equation}

The contravariant components of the metric tensor can be written in terms of a null tetrad basis $\mathbf{e}^{a}=(\mathbf{l},\mathbf{n},\mathbf{m},\mathbf{\bar{m}})$, as follows:
\begin{equation}
\label{frame_metric}g^{\mu\,\nu}=-l^\mu\,n^\nu-l^\nu\,n^\mu+m^\mu\,\bar{m}^\nu+m^\nu\,\bar{m}^\mu,
\end{equation}
where
$l_\mu\,l^\mu=n_\mu\,n^\mu=m_\mu\,m^\mu=l_\mu\,m^\mu=n_\mu\,m^\mu=0$. 
Moreover, $l_\mu\,n^\mu=-m_\mu\,\bar{m}^\mu=-1$.
For the line element \eqref{EDF_coord}, we can make the specific choice
\begin{align}
\label{tetrad0}&l^\mu=\delta^\mu_r,\\
&n^\mu=\sqrt{\frac{F}{G}}\delta^\mu_u-\frac{F}{2}\delta^\mu_r,\\
&m^\mu=\frac{1}{\sqrt{2\,H}}\left(\delta^\mu_\theta+\frac{i}{\sin\theta}\,\delta^\mu_\phi\right).
\end{align}

Performing a complex transformation of coordinates, given by
\begin{align}
&r'= r+i\,a\,\cos\theta,\\
&u'= u -i\,a\,\cos\theta,\\
&\theta'=\theta,\\
&\phi'=\phi, 
\end{align}
the null tetrad basis vectors, in the primed coordinate system, become 
\begin{align}
\label{lprime}&l'^\mu=\delta^\mu_{r'}\\
&n'^\mu=\sqrt{\frac{B}{A}}\delta^\mu_{u'}-\frac{B}{2}\delta^\mu_{r'}\\
\label{mprime} &m'^\mu=\frac{1}{\sqrt{2\Psi}}\left[\delta^\mu_{\theta'}+i\,a\,\sin\theta'\left(\delta^\mu_{u'}-\delta^\mu_{r'}\right)+\frac{i}{\sin\theta'}\,\delta^\mu_{\phi'}\right].
\end{align}
In the latter the functions $G(r),F(r)$ and $H(r)$ were replaced as:
\begin{align}
\label{GtoA}&G(r)\rightarrow A(r',\theta',a),\\
\label{FtoB}&F(r)\rightarrow B(r',\theta',a),\\
\label{HtoPsi}&H(r)\rightarrow \Psi(r',\theta',a),
\end{align}
where A, B and $\Psi$ are real functions that will be fixed later. In order to recover the spherically symmetric seed metric in the non-rotating case, it is imposed that
\begin{align}
\label{a-goes-to-zero}
\lim_{a\rightarrow 0}A(r', \theta', a)= G(r'),\\
\lim_{a\rightarrow 0}B(r', \theta', a)= F(r'),\\
\lim_{a\rightarrow 0}\Psi(r', \theta', a)= H(r').
\label{a-goes-to-zero2}
\end{align}

The modification on the original NJA arises in Eqs.~\eqref{GtoA}-\eqref{HtoPsi}, where the functions A, B, and $\Psi$ would be determined by the complexification of the radial coordinate $r$. In the MNJA, {A, B, $\Psi$} are fixed by using another criteria and by requiring a particular form for the stress-energy tensor.

Using Eqs.~\eqref{frame_metric} and \eqref{lprime}-\eqref{mprime}, we find that
\begin{align}
&g^{uu}=\frac{a^2\,\sin^2\theta}{\Psi}, \quad g^{ur}=-\sqrt{\frac{B}{A}}-\frac{a^2\,\sin^2\theta}{\Psi},\\
&g^{u\phi}=\frac{a}{\Psi}, \quad g^{rr}=B+\frac{a^2\,\sin^2\theta}{\Psi},\\
&g^{r\phi}=-\frac{a}{\Psi},\quad g^{\theta\theta}=\frac{1}{\Psi}, \quad g^{\phi\phi}=\frac{1}{\Psi\,\sin^2\theta},
\end{align}
where we dropped the primes for convenience. The corresponding line element in the advanced null coordinates is
\begin{align}
\nonumber &ds^2=-A\,du^2-2\,\sqrt{\frac{A}{B}}\,du\,dr+\Psi\,d\theta^2\\
\nonumber &-2\,a\,\sin^2\theta\left(\sqrt{\frac{A}{B}}-A\right)du\,d\phi+2\,a\,\sin^2\theta\,\sqrt{\frac{A}{B}}dr\,d\phi\\
\label{lineelEF} &+\sin^2\theta\left[\Psi+a^2\,\sin^2\theta\left(2\,\sqrt{\frac{A}{B}}-A\right)\right]\,d\phi^2.
\end{align}

The final step is to write the line element \eqref{lineelEF} in Boyer-Lindquist-like coordinates. This is accomplished by the following coordinate transformation:
\begin{align}
\label{BL1}du=dt+\chi_1(r)\,dr,\\
\label{BL2}d\phi=d\varphi+\chi_2(r)\,dr,
\end{align}
and imposing that $g_{r\varphi}$ and $g_{tr}$ are zero. This is not always possible for the NJA, since the functions $\chi_1(r)$ and $\chi_2(r)$ may depend on $\theta$. If so, the right hand side of Eqs.~\eqref{BL1}-\eqref{BL2} is not an exact differential and thus no coordinates $u(t,r)$ and $\phi(\varphi,r)$ exist.~\footnote{It was recently shown that a rotating solution obtained from the NJA only admits separability {for null geodesics} if $\chi_1$ and $\chi_2$~are functions of the radial coordinate only, what implies that the metric can be written in Boyer-Lindquist coordinates~\cite{NJA_shadow}.} This (potential) failure of the NJA to produce a circular metric is related to the fact that the functions {$A$, $B$, $\Psi$} are fixed by the complexification of the radial coordinate $r$. In the MNJA, on the other hand, the functions $A$, $B$ and $\Psi$ are still not fixed, $i.e.$ they are unknowns. Therefore, if
\begin{align}
&\chi_1=-\frac{K+a^2}{F\,H+a^2},\\
&\chi_2=-\frac{a}{F\,H+a^2},
\end{align}
where
\begin{equation}
\label{K(r)}K(r)=\sqrt{\frac{F(r)}{G(r)}}\,H(r),
\end{equation}
we can always write the line element~\eqref{lineelEF} 
in Boyer-Lindquist-like coordinates, provided that
\begin{align}
&A(r,\theta)=\frac{\left(F\,H+a^2\,\cos^2\theta\right)}{\left(K+a^2\,\cos^2\theta\right)^2}\,\Psi,\\
&B(r,\theta)=\frac{F\,H+a^2\,\cos^2\theta}{\Psi}.
\end{align}
We remark that the latter satisfies the requirement of Eqs.~(\ref{a-goes-to-zero})-(\ref{a-goes-to-zero2}).
Then, the new geometry, written in Boyer-Lindquist-like coordinates, is given by
\begin{align}
\nonumber &ds^2=-\frac{\left(F\,H+a^2\,\cos^2\theta\right)}{\left(K+a^2\,\cos^2\theta\right)^2}\Psi\,dt^2\\
\nonumber &-2\,a\,\sin^2\theta\left[\frac{K-F\,H}{\left(K+a^2\,\cos^2\theta\right)^2}\right]\,\Psi\,dt\,d\varphi +\frac{\Psi}{F\,H+a^2}\,dr^2\\
\label{Rot_sol}&+\Psi\,d\theta^2\\
\nonumber &+\Psi\,\sin^2\theta\left[1+a^2\,\sin^2\theta\,\frac{2\,K-F\,H+a^2\,\cos^2\theta}{\left(K+a^2\,\cos^2\theta\right)^2}\right]\,d\varphi^2.
\end{align}

We note that the function $\Psi(r,\theta,a)$ is still 
an unfixed
function present in the line element \eqref{Rot_sol}. It is possible to determine $\Psi(r,\theta,a)$ by imposing some constraint in the stress-energy tensor $T^{\mu\,\nu}$. As pointed out in Ref.~\cite{NJA_mod2}, if the source $T^{\mu\,\nu}$ represents an imperfect fluid rotating about the $z$-axis, $\Psi$ obeys the following nonlinear differential equations~\cite{NJA_mod2}:
 \begin{align}
\label{Psi_eq_1}\left(K+a^2\,y^2\right)^2\left(3\,\Psi_{,r}\,\Psi_{,y^2}-2\,\Psi\,\Psi_{,r y^2}\right)=3\,a^2\,K_{,r}\,\Psi^2,\\
\nonumber \left[K_{,r}^2+K\,\left(2-K_{,r r}\right)-a^2\,y^2\left(2+K_{,r r}\right)\right]\,\Psi \ \ \ \ \ \ \ \ \ \ \ \ \\
\label{Psi_eq_2}+\left(K+a^2\,y^2\right)\left(4\,y^2\,\Psi_{,y^2}-K_{,r}\Psi_{,r}\right)=0,\ \ \ \ 
\end{align}
where $y\equiv \cos\theta$, and the comma in the subscript denotes differentiation with respect to $r$ or/and $y$. 
It may be observed, however, that $\Psi$ is an overall conformal factor in~\eqref{Rot_sol}. Thus, neither the causal structure nor the null geodesic flow will depend on the choice of $\Psi$.

There are several examples in the literature where the 
MNJA 
was used in order to generate rotating solutions in Boyer-Lindquist-like coordinates (cf., for instance, Refs.~\cite{MNJ_Sol_I,MNJ_Sol_II,MNJ_Sol_III,MNJ_Sol_IV,MNJ_Sol_V,MNJ_Sol_VI,MNJ_Sol_VII,MNJ_Sol_VIII}).
%

\section{Hamilton-Jacobi equation for null geodesics in rotating spacetime obtained through modified Newman-Janis algorithm}
\label{SecIII}
Recently, it was shown that if the NJA succeeds in bringing the rotating solution to the Boyer-Lindquist-like coordinates, the Hamilton-Jacobi equation admit separability \cite{NJA_shadow}.  Moreover, Azreg-A{\"inou} studied the separability of the Hamilton-Jacobi equation for regular black holes with $F(r)=G(r)$ and $H(r)=r^2$~\cite{NJA_mod}. One may ask if a \textit{generic} rotating solution obtained through the 
MNJA 
also admits separability 
of 
the Hamilton-Jacobi equation.  We note that due to the complicated form of \eqref{Psi_eq_1} and \eqref{Psi_eq_2}, only few solutions for $\Psi$ are known~\cite{NJA_mod}. Most of them do not correspond to \eqref{K(r)} with $F(r)\neq G(r)$. However, as we show below, for null geodesics, we do not need to determine the explicit form of $\Psi$, since the separability of the Hamilton-Jacobi equation is independent of this function.

We use the Hamilton-Jacobi equation, given by
\begin{equation}
\label{HJ_eq}\frac{\partial S}{\partial \tau}+H=0,
\end{equation}
where $S$ is the Jacobi action, $\tau$ is an affine parameter, and $H$ is the Hamiltonian 
\begin{equation}
\label{Hamiltonian}H=\frac{1}{2}\,g^{\mu\,\nu}p_{\mu}\,p_{\nu}.
\end{equation}
The relation between the Jacobi action and the momentum $p^\mu$ is
\begin{equation}
\frac{\partial S}{\partial x^\mu}=p_\mu.
\end{equation}

Since the metric tensor does not depend on $t$ and $\varphi$, we have two conserved quantities, $E=-p_t$ and $p_\varphi=\Phi$, which are the energy and angular momentum of the photon with respect to the axis of symmetry, respectively. We assume the following Ansatz for the Jacobi action $S$:
\begin{equation}
\label{Ansatz_S}S=\frac{\mu^2}{2}\,\tau-E\,t+\Phi\,\varphi+S_r(r)+S_\theta(\theta),
\end{equation}
where $\mu$ is the mass of the particle. Using the Ansatz \eqref{Ansatz_S} in Eq.~\eqref{HJ_eq}, we find that
\begin{align}
\label{HJ_eq2}&-\frac{1}{\Psi\left(F\,H+a^2\right)}\,\left[E\,\left(K+a^2\right)-a\,\Phi\right]^2+\frac{1}{\Psi}\left(p_\theta\right)^2\\
\nonumber& +\frac{1}{\Psi\,\sin^2\theta}\left(a\,E\,\sin^2\theta-\Phi\right)^2+\frac{F\,H+a^2}{\Psi}\left(p_r\right)^2=-\mu^2.
\end{align}
In the particular case of null geodesics ($\mu=0$) the unfixed function $\Psi$ disappears, and we can rewrite Eq.~\eqref{HJ_eq2} as 
\begin{align}
\nonumber&\left(F\,H+a^2\right)\,\left(p_r\right)^2-\frac{1}{\left(F\,H+a^2\right)}\left[E\,\left(K+a^2\right)-a\,\Phi\right]^2=\\ 
\label{HJ_eq3}&-\left[ \left(p_\theta\right)^2+\frac{1}{\sin^2\theta}\left(a\,E\,\sin^2\theta-\Phi\right)^2\right]\equiv -\mathcal{K}.
\end{align}
We point out that $F$, $H$ and $K$ are functions of the radial coordinate $r$ only, therefore the right-hand side of Eq.~\eqref{HJ_eq3} is function of $r$ only, while the left-hand side is a function of $\theta$ only. This equality will only hold if both sides are equal to a constant $\mathcal{K}$. We rewrite $\mathcal{K}$ as
\begin{equation}
\mathcal{K}\equiv Q+\left(a\,E-\Phi\right)^2,
\end{equation} 
where $Q$ is called Carter's constant. Then, we find from Eq.~\eqref{HJ_eq3}, that the equations for $p_r$ and $p_\theta$ are separable:
\begin{align}
\label{pr}\nonumber &\left(F\,H+a^2\right)^2\left(p_r\right)^2=\left[E\,\left(K+a^2\right)-a\,\Phi\right]^2
\\ &-\left(F\,H+a^2\right)\,\left[Q+\left(a\,E-\Phi\right)^2\right],\\
\label{ptheta}&\left(p_\theta\right)^2=Q+a^2\,E^2\,\cos^2\theta-\frac{\cos^ 2\theta}{\sin^2\theta}\,\Phi^2.
\end{align}

From Eqs.~\eqref{pr}-\eqref{ptheta}, we see that the Hamilton-Jacobi equation (for null geodesics) is completely separable for the general line element~\eqref{Rot_sol}. Let us emphasize this point: the Newman-Janis algorithm admits separability only if the generated rotating spacetime can be written in Boyer-Lindquist-like coordinates; by contrast, the 
MNJA 
always admits separability in the Hamilton-Jacobi equation, for null geodesics. In addition, since the complexification of the radial coordinate is introduced in a different fashion, the resulting equations of motion will be related to the functions present in the spherically symmetric seed metric, {\it i.e.} 
$F(r)$, $H(r)$ and $K(r)$.

\section{Hamilton-Jacobi equation in rotating spacetime obtained through modified Newman-Janis algorithm in plasma}
\label{SecPlasma}
{In this section, we study the separability of Hamilton-Jacobi equation in a rotating spacetime obtained through MNJA in the presence of a plasma. We consider a cold, pressureless and non-magnetized plasma model around the black hole. In the presence of a plasma, the light rays do not move along null geodesics. The propagation of light rays in the plasma is described  by the following Hamiltonian \cite{Perlick_book,Synge_book}:
\begin{equation}
\label{Hamiltonianp}H_p=\frac{1}{2}\left(g^{\mu\,\nu}\,p_{\mu}\,p_{\nu}+\omega_p^2(r,\theta)\right),
\end{equation}
where $\omega_p$ is the plasma electron frequency. We point out that the difference between null geodesic [Eq.~\eqref{Hamiltonian}] and the plasma case [Eq.~\eqref{Hamiltonianp}] is the term $\omega_p^2$ in Eq.~(\ref{Hamiltonianp}). The plasma frequency $\omega_p$ is related to the electron density $N_e$ by
\begin{equation}
\omega_p(r,\theta)=\frac{4\,\pi\,e^2}{m_e}\,N_e(r,\theta),
\end{equation}
where $e$ and $m_e$ are the charge and mass of the electron, respectively. For the Kerr geometry, it was shown that the Hamilton-Jacobi equation for the Hamiltonian $H_{p}$ is separable if the function $\omega_p^2$ has a particular form \cite{Kerr_Plasma}. One can ask if the Hamilton-Jacobi equation for the Hamiltonian~\eqref{Hamiltonianp} is separable in a generic spacetime obtained through the MNJA. In order to answer this question, we start applying the Hamilton-Jacobi equation~\eqref{HJ_eq} to the Hamiltonian~\eqref{Hamiltonianp}, using that the Jacobi action $S$ is given by Eq.~\eqref{Ansatz_S}. By substituting Eq.~\eqref{Ansatz_S} in Eq.~\eqref{HJ_eq}, and using the Hamiltonian given in \eqref{Hamiltonianp}, we find that
\begin{align}
\nonumber \left(a\,\sin\theta\,E-\frac{\Phi}{\sin\theta} \right)^2-\frac{1}{\left(F\,H+a^2\right)}\left[\left(K+a^2\right)\,E-a\,\Phi\right]^2\\
\label{HJP_eq1}+\left(F\,H+a^2\right)\,\left(p_r\right)^2+\left(p_\theta\right)^2+\Psi\,\omega_p^2(r,\theta)=0.
\end{align}} 

{From Eq.~\eqref{HJP_eq1}, we see that the Hamilton-Jacobi equation is not separable in the general case of light rays propagating in a plasma, since the last term on the left-hand side  is an arbitrary function of $r$ and $\theta$. However, if
\begin{equation}
\label{Sep_cond_plasma}\omega_p^2(r,\theta)=\frac{\omega_r(r)+\omega_\theta(\theta)}{\Psi(r,\theta)},
\end{equation}
the Hamilton-Jacobi equation for light rays in the presence of a plasma is separable as
\begin{align}
\nonumber &\left(F\,H+a^2\right)^2\,\left(p_r\right)^2=\left[\left(K+a^2\right)\,E-a\,\Phi\right]^2\\
\label{pr_p}&-\left(F\,H+a^2\right)\left[Q_p+(a\,E-\Phi)^2+\omega_r\right],\\
\label{ptheta_p}&\left(p_\theta\right)^2=Q_p-\omega_\theta+\cos^2\theta\left(a^2\,E^2-\frac{\Phi^2}{\sin^2\theta}\right),
\end{align}
where $Q_p$ is a generalized Carter-like constant, related to the propagation of light rays in the presence of the plasma obeying Eq.~(\ref{Sep_cond_plasma}). We note that Eq.~\eqref{pr_p} depends only on the radial coordinate $r$, while Eq.~\eqref{ptheta_p} depends only on $\theta$, as expected from the separability of the Hamilton-Jacobi equation. Moreover, if $\omega_r=\omega_\theta=0$, we recover the null geodesic results presented in Sec.~\ref{SecIII}. }

{Hence, Eq.~\eqref{Sep_cond_plasma} establishes the condition for the separability of the Hamilton-Jacobi equation for the propagation of light rays in the presence of a plasma. The analogous result for Kerr spacetime was obtained in Ref.~\cite{Kerr_Plasma}, with the choice $\Psi(r,\theta)=r^2+a^2\,\cos^2\theta$.}
{If the plasma condition ~\eqref{Sep_cond_plasma} is not satisfied, the equations of motion may stop being separable.}


\section{Shadows}
\label{SecIV}
In this section, we study the analytical form of the shadow cast by rotating black holes obtained through the MNJA, as an application of the separability properties of null geodesics studied in Sec.~\ref{SecIII}. In particular,  we study the spherical photon orbits, which are related to the shadows cast by the general line element \eqref{Rot_sol}. We also reproduce the shadows of specific solutions presented in the literature. 

We point out that Eqs.~\eqref{pr}-\eqref{ptheta} can be rewritten as
\begin{align}
\nonumber &\Psi^2\,\left(\dot{r}\right)^2=\left[E\,\left(K+a^2\right)-a\,\Phi\right]^2\\
\label{rdot}&-\left(F\,H+a^2\right)\left[Q+\left(a\,E-\Phi\right)^2\right] \\
\label{thetadot}&\Psi^2\,\left(\dot{\theta}\right)^2=Q+\cos^2\theta\left(a^2\,E^2-\frac{\Phi^2}{\sin^2\theta}\right)
\end{align}
where we have used that $p^\mu=g^{\mu\,\nu}\,p_\nu=\dot{x}^\mu$, and $\dot{x}^\nu=dx^\nu/d\tau$. The equations of motion for $(t,\varphi)$ are given by
\begin{align}
\nonumber &\dot{t}=g^{t\,\mu}\,p_{\mu}=\frac{\left(K+a^2\right)^2-a^2\,\sin^2\theta\,\left(F\,H+a^2\right)}{\Psi\,\left(F\,H+a^2\right)}E\\
&+\frac{a\,\left(F\,H-K\right)}{\Psi\left(F\,H+a^2\right)}\,\Phi,\\
\nonumber &\dot{\varphi}=g^{\varphi\,\mu}\,p_{\mu}=\frac{\left(F\,H+a^2\,\cos^2\theta\right)}{\Psi\,\sin^2\theta\left(F\,H+a^2\right)}\Phi\\
\label{phidot}&-\frac{a\,\left(F\,H-K\right)}{\Psi\left(F\,H+a^2\right)}\,E.
\end{align}

We define the functions $\mathcal{R}(r)$ and $\Theta(\theta)$, such that
\begin{align}
\label{mathcalR}&\frac{\Psi^2\,\left(\dot{r}\right)^2}{E^2}=\mathcal{R}(r),\\
\label{mathcalTheta}&\frac{\Psi^2\,\left(\dot{\theta}\right)^2}{E^2}=\Theta(\theta),
\end{align}
where
\begin{align}
\nonumber &\mathcal{R}(r)=\left(K+a^2-a\,\lambda\right)^2\\
\label{R}&-\left(F\,H+a^2\right)\,\left[\eta+\left(a-\lambda\right)^2\right],\\
&\Theta(\theta)=\eta+\cos^2\theta\left(a^2-\frac{\lambda^2}{\sin^2\theta}\right).
\end{align}

We point out that $\mathcal{R}(r)\geqslant 0$ and $\Theta(\theta)\geqslant 0$, since the left-hand side of Eqs.~\eqref{mathcalR}-\eqref{mathcalTheta} is always positive. In addition, we have introduced the constants $\eta$ and $\lambda$:
\begin{align}
&\eta=\frac{Q}{E^2},\\
&\lambda=\frac{\Phi}{E}.
\end{align}

{In a separable coordinate chart, the Spherical Photon Orbits are a set of light ray orbits with a constant radial coordinate $r$. These orbits can be computed by solving simultaneously}
\begin{align}
\label{SPO_1}&\mathcal{R}=0,\\
\label{SPO_2}&\frac{d \mathcal{R}}{d\,r}=0.
\end{align}

Substituting Eq.~\eqref{R} into Eqs.~\eqref{SPO_1} and \eqref{SPO_2}, we find two solutions for $\lambda$, namely 
\begin{align}
&\lambda_{I}=\frac{K+a^2}{a},\\
\label{lambda_SPO}&\lambda_{II}=\frac{K+a^2}{a}-\frac{2\,K'}{a}\,\frac{\left(F\,H+a^2\right)}{\left(H\,F\right)'},
\end{align}
where the primes denote derivative with respect to $r$. The spherical photon orbits are described by $\lambda_{II}$, given in Eq.~\eqref{lambda_SPO}. We can find the corresponding value for $\eta_{II}$ by substituting $\lambda_{II}$ into \eqref{SPO_1} or \eqref{SPO_2}, obtaining 
\begin{align}
\nonumber \eta_{II}=&\frac{4\,\left(a^2+F\,H\right)}{\left(H\,F\right)'^2}\,K'^2\\
\label{eta_SPO}&-\frac{1}{a^2}\left[K-\frac{2\,\left(F\,H+a^2\right)}{\left(H\,F\right)'}\,K'\right]^2.
\end{align}

Equations \eqref{lambda_SPO} and \eqref{eta_SPO} are evaluated on a given radius $r_{\textrm{SPO}}$, which is the radius of the spherical photon orbits. We recall that $\eta$ and $\lambda$ are constants of motion. Equations~\eqref{lambda_SPO} and \eqref{eta_SPO} are the values of such constants that ensure the existence of a spherical photon orbit with radius $r_{\textrm{SPO}}$. However, the physical range of $r_{\textrm{SPO}}$ is restricted by the motion in $\theta$, $i.e.$ $p_\theta^2\geqslant 0$. 
The edge of the shadow is determined by the unstable spherical photon orbits. For observers far away from the black hole, the coordinates of the shadow edge, in a plane perpendicular to the line joining the observer and the origin of the radial coordinate, are given by \cite{Shadow_Kerr_BHS,Chandra} 
\begin{align}
\label{x'_0}&x'=\lim_{r_0\rightarrow\infty}\left(-r_0^2\,\sin\theta_0\,\left.\frac{d\varphi}{dr}\right|_{(r_0,\theta_0)}\right),\\
\label{y'_0}&y'=\lim_{r_0\rightarrow\infty}\left(r_0^2\,\left.\frac{d\theta}{dr}\right|_{(r_0,\theta_0)}\right),
\end{align}
where $r_0$, $\theta_0$ are the $r$ and $\theta$ coordinates of the observer. Assuming that both the static seed metric and the generated rotating geometry are asymptotic flat, we have $G(r)\rightarrow 1$, $F(r)\rightarrow 1$, $H(r)\rightarrow r^2$, $\Psi\rightarrow r ^2$ when $r\rightarrow \infty$. Hence, inserting Eqs.~\eqref{rdot},  \eqref{thetadot}, \eqref{phidot} into Eqs.~\eqref{x'_0} and \eqref{y'_0}, we find that
\begin{align}
\label{x'}&x'=-\frac{\lambda_{II}}{\sin\theta_0},\\
\label{y'}&y'=\pm \sqrt{\eta_{II}+a^2\,\cos^2\theta_0-\frac{\lambda_{II}^2}{\tan^2\theta_0}}.
\end{align} 
The shadow edge is obtained by using the values of $\eta_{II}$ and $\lambda_{II}$, given in Eqs.~\eqref{lambda_SPO}- \eqref{eta_SPO}, into Eqs.~\eqref{x'} and \eqref{y'}.

\subsection{Shadow results: Generic magnetically charged regular black hole solution}
In this subsection, we study the shadow of a particular rotating solution, in order to show the applicability of the results presented in Sec.~\ref{SecIII} and \ref{SecIV}. Our seed metric is the spherically symmetric family of generic magnetically charged regular black hole spacetimes, proposed in Ref.~\cite{R_BH} by Fan and Wang. The line element is given by:
\begin{equation}
\label{RBH_SOL}ds^2=-f(r)\,dt^2+\frac{1}{f(r)}\,dr^2+r^2\,d\Omega^2,
\end{equation}
where $d\Omega^2=d\theta^2+\sin^2\theta\,d\phi^2$ is the line element of the unit 2-sphere, and
\begin{equation}
f(r)=1-\frac{2\,M\,r^{\epsilon-1}}{\left(r^\kappa+g^\kappa\right)^\frac{\epsilon}{\kappa}},
\end{equation}
with $M=q^3/\alpha$ being the pure gravitational mass, according to Refs.~\cite{R_BH_1,R_BH_2}.

The parameter $\epsilon$ must be greater or equal to 3 \cite{R_BH,R_BH_1,R_BH_2} and characterizes the degree of non-linearity, while $\alpha$ is related to the strength of the nonlinear effects. $g$ and $\nu$ are free parameters related to the magnetic charge and the character of its field, respectively. 

Based on this generic black hole spacetime, we can describe several known regular black hole solutions, for instance:
\begin{enumerate}[label=(\roman*)]
\item $\kappa=2$ and $\epsilon=3$ represents the Bardeen solution:
\begin{equation}
 f(r)=1-\frac{2\,M\,r^2}{\left(r^2+g^2\right)^\frac{3}{2}}.
 \end{equation}
 \item $\kappa=3$ and $\epsilon=3$ represents the Hayward solution:
 \begin{equation}
 f(r)=1-\frac{2\,M\,r^2}{\left(r^3+g^3\right)}.
 \end{equation}
 \item  For $\kappa=1$, we have a different class of regular black hole solutions:
 \begin{equation}
 f(r)=1-\frac{2\,M\,r^{\epsilon-1}}{\left(r+g\right)^\epsilon}.
 \end{equation}
 
\end{enumerate} 

The rotating generalization of the line element~\eqref{RBH_SOL} and some of its properties were studied in Ref.~\cite{MNJ_Sol_I}. 

Let us now study the shadows cast by these rotating solutions.  Comparing Eqs.~\eqref{RBH_SOL} and \eqref{linel}, we find that
\begin{align}
&G(r)=F(r)=1-\frac{2\,M\,r^{\epsilon-1}}{\left(r^\kappa+g^\kappa\right)^\frac{\epsilon}{\kappa}},\\
&H(r)=r^2,\\
&K(r)=r^2.
\end{align}
Moreover, it is straightforward to check that 
\begin{equation}
\Psi(r,\theta)=r^2+a^2\,\cos^2\theta
\end{equation}
is a solution to Eqs.~\eqref{Psi_eq_1}-\eqref{Psi_eq_2}. Then, the line element for the rotating generic magnetically charged regular black hole is given by
\begin{align}
\nonumber ds^2=-\left(1-\frac{2\,m(r)\,r}{\Psi}\right)\,dt^2-\frac{4\,m(r)\,r\,a\sin^2\theta}{\Psi}\,dt\,d\varphi+\frac{\Psi}{\Delta}\,dr^2\\
\label{Rot_RBH}+\Psi\,d\theta^2+\sin^2\theta\left[\left(r^2+a^2+b^2\right)+\frac{2\,m(r)\,r\,a^2\,\sin^2\theta}{\Psi}\right]\,d\varphi^2,
\end{align}  
where
\begin{align}
&m(r)=\frac{M\,r^{\epsilon}}{\left(r^\kappa+g^\kappa\right)^\frac{\epsilon}{\kappa}},\\
&\Delta=r^2-2\,m(r)\,r+a^2.
\end{align}

\begin{figure*}
  \centering
  \subfigure{\includegraphics[scale=0.8]{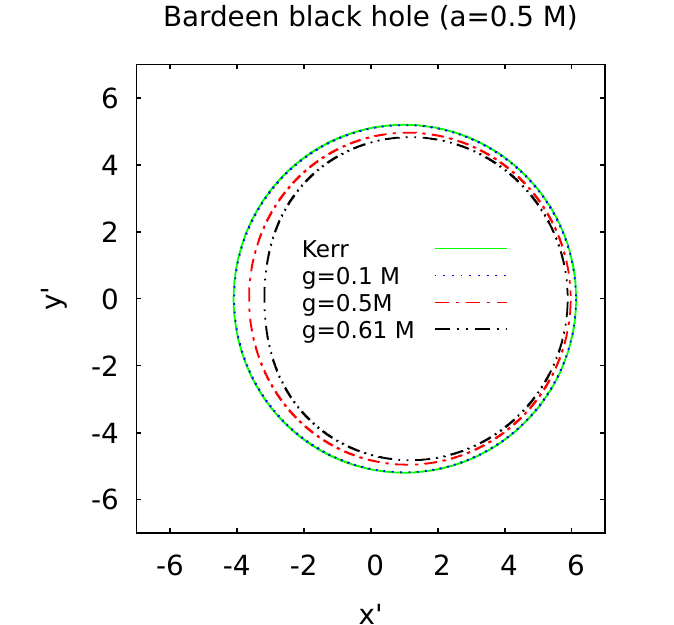}}
  \subfigure{\includegraphics[scale=0.8]{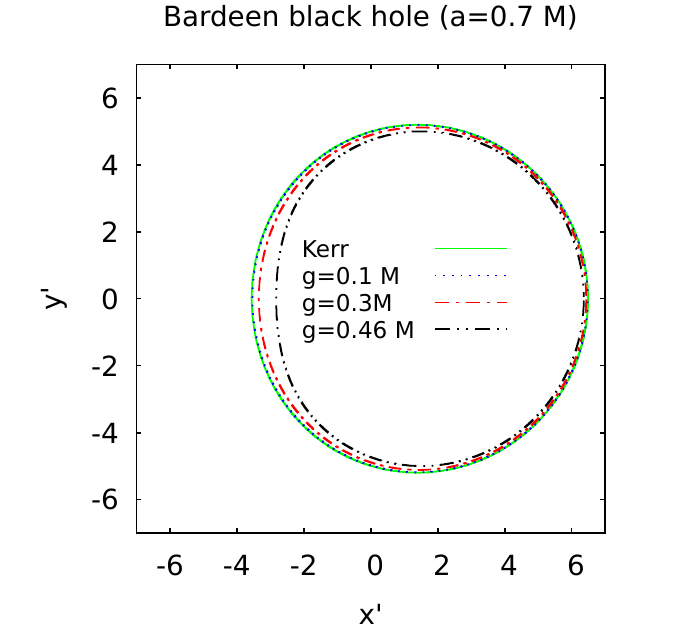}}
\subfigure{\includegraphics[scale=0.8]{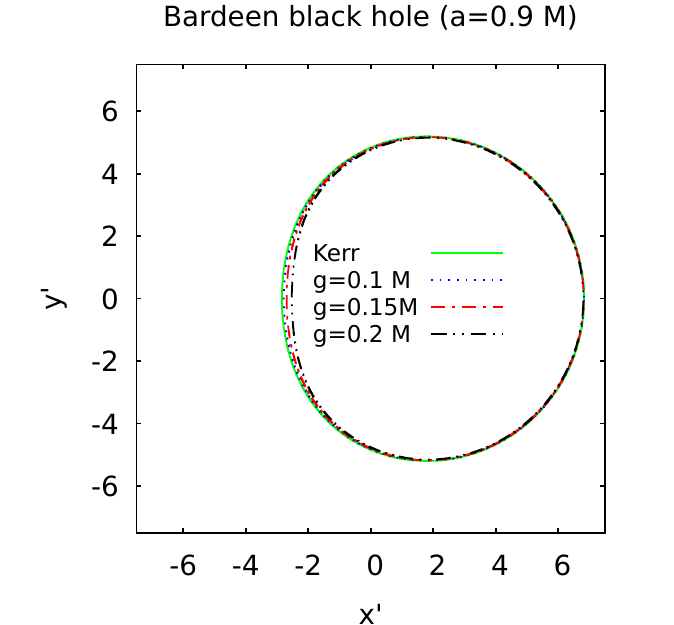}}
\quad
\subfigure{\includegraphics[scale=0.8]{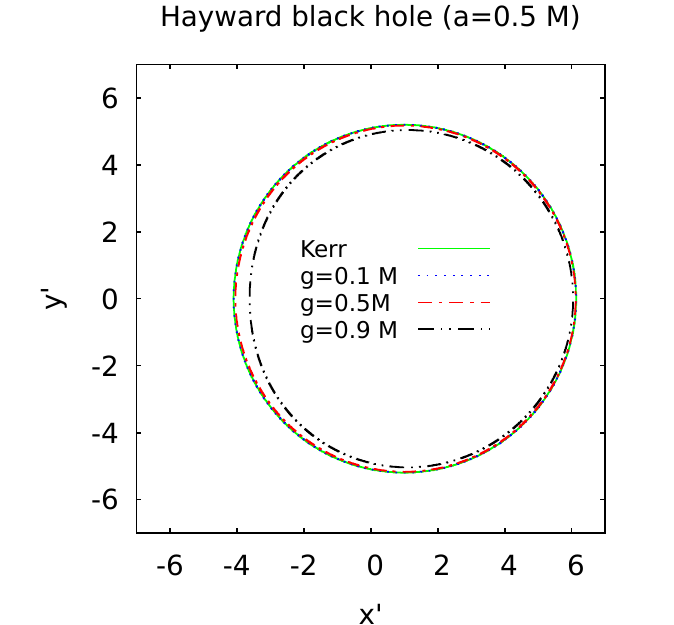}}
\subfigure{\includegraphics[scale=0.8]{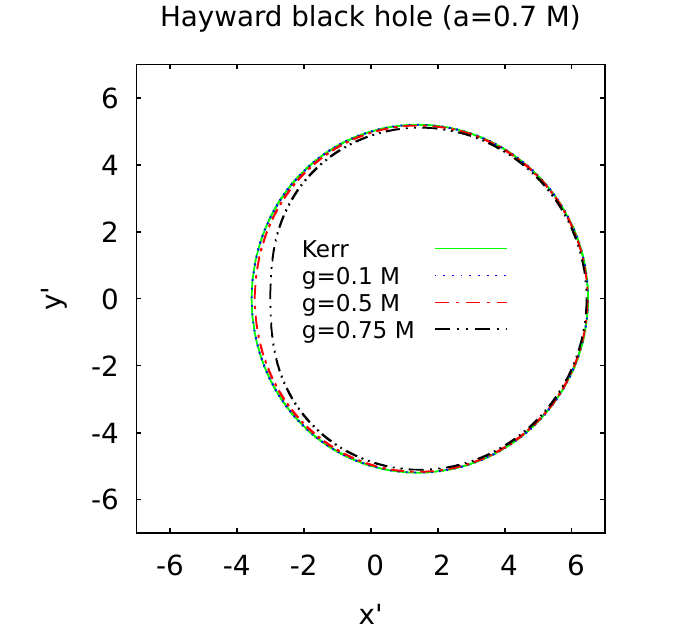}}
\subfigure{\includegraphics[scale=0.8]{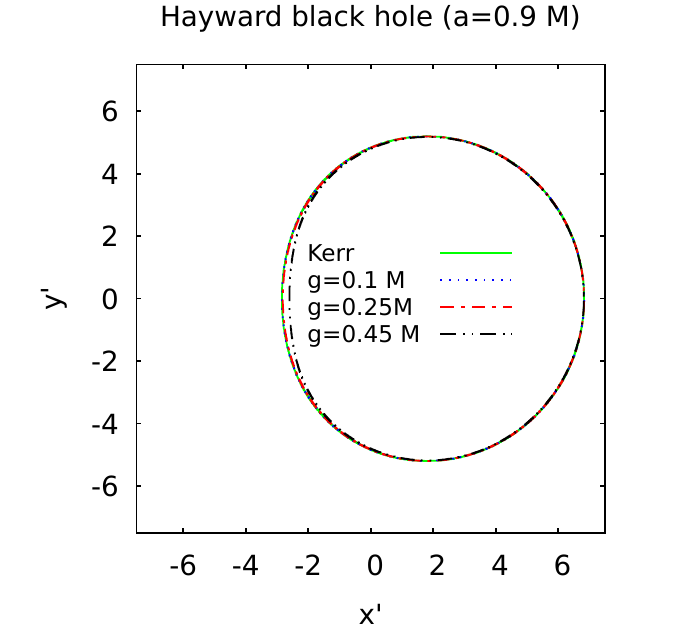}}
\caption{Top row: Shadows of rotating Bardeen black holes for different values of magnetic charge and rotation parameter. Bottom row: Shadows of rotating Hayward black holes for different values of magnetic charge and rotation parameter. The shadows of the Bardeen and Hayward are compared to the shadow of an uncharged Kerr black hole.} 
\label{Fig1}
\end{figure*}

We can obtain the shadows of generic regular black holes through Eqs.~\eqref{x'} and \eqref{y'}. The results for the Bardeen ($\epsilon=3$, $\kappa=2$) and Hayward ($\epsilon=3$, $\kappa=3$) black holes are given in Fig.~\ref{Fig1}. Such results are known in the literature and were obtained, for instance, in Ref.~\cite{R_BH_3}. 

We may also obtain the shadows of a different class of regular black holes by choosing different values for $\epsilon$ and $\kappa$, for instance, $\epsilon\geqslant 3$ and $\kappa=1$. Such regular solution approaches a  Maxwellian field in the weak field limit \cite{R_BH}.

\begin{figure*}
  \centering
  \subfigure{\includegraphics[scale=0.8]{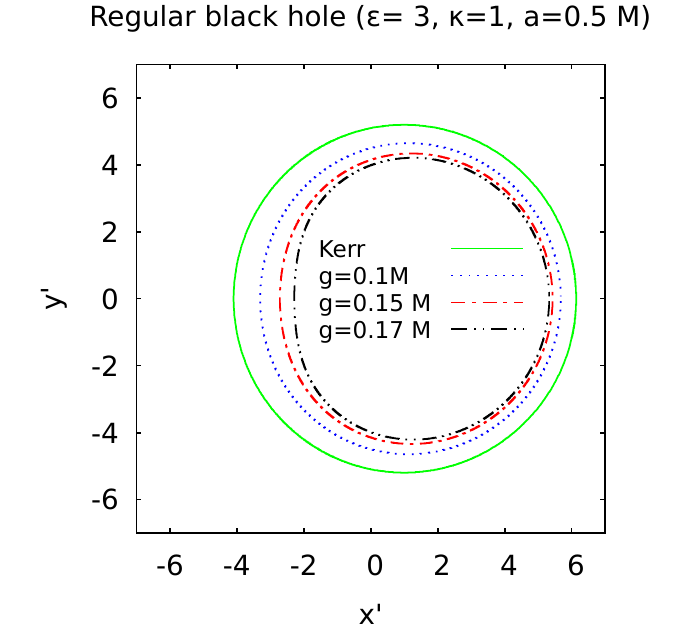}}
  \subfigure{\includegraphics[scale=0.8]{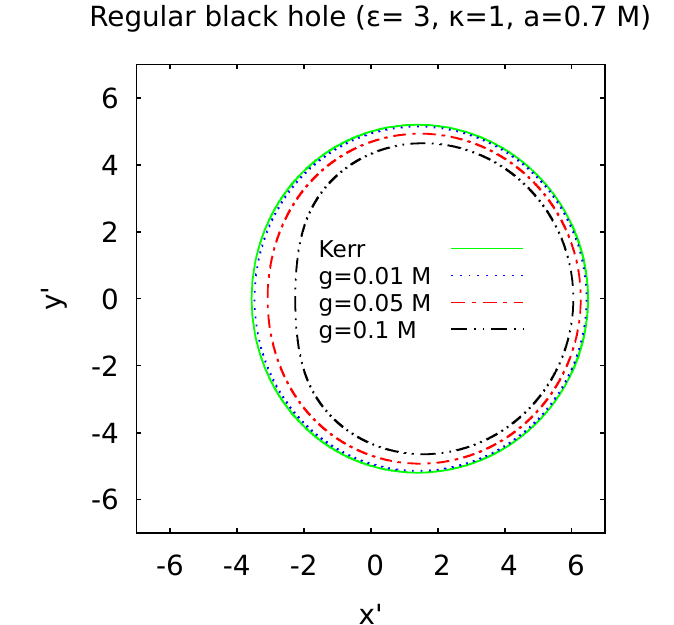}}
\subfigure{\includegraphics[scale=0.8]{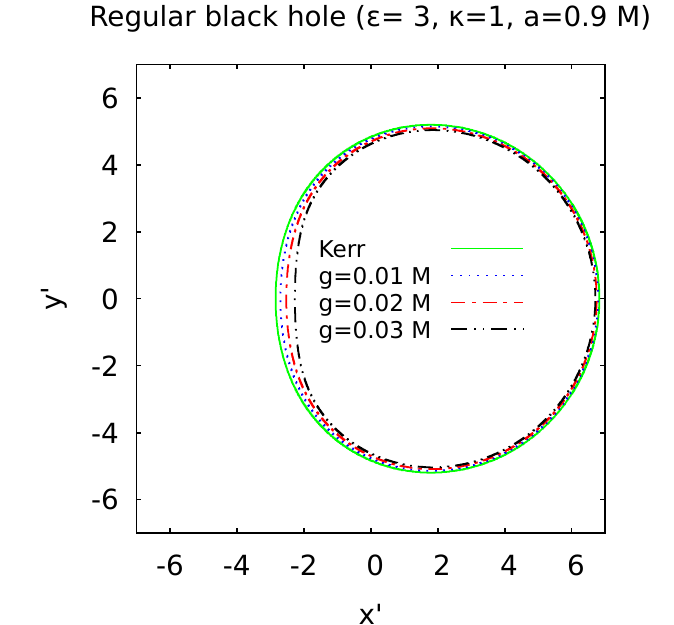}}
\qquad
  \subfigure{\includegraphics[scale=0.8]{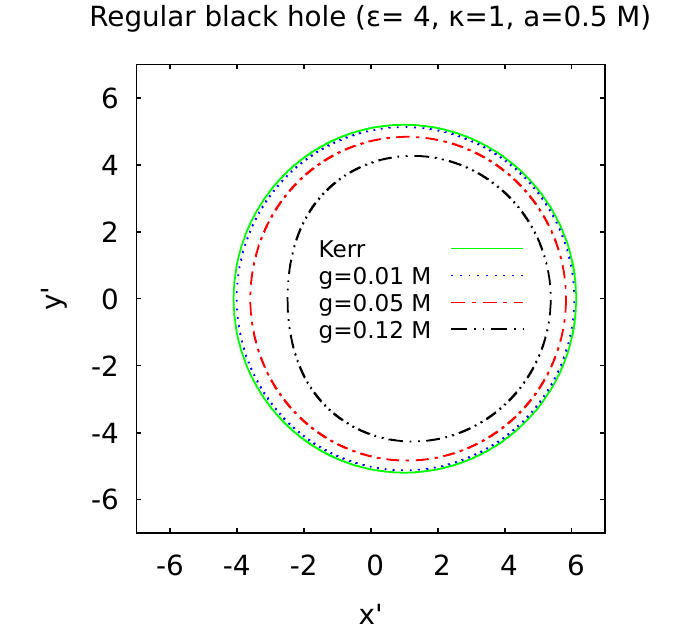}}
  \subfigure{\includegraphics[scale=0.8]{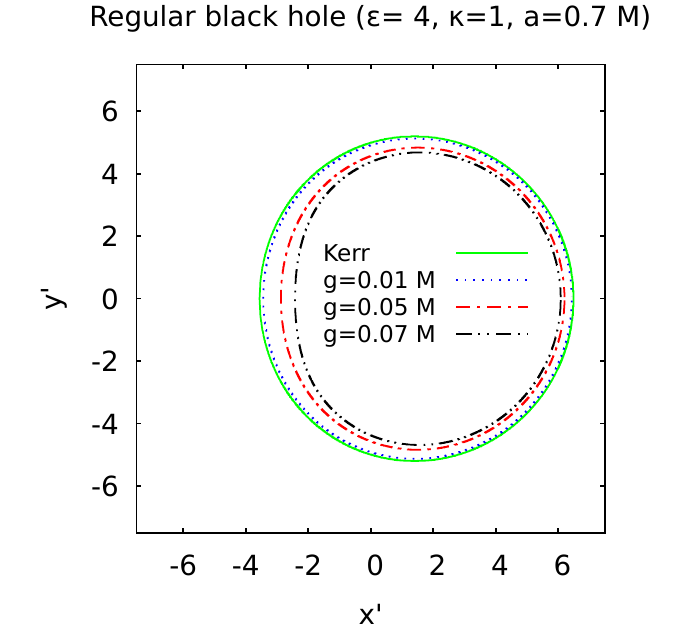}}
\subfigure{\includegraphics[scale=0.8]{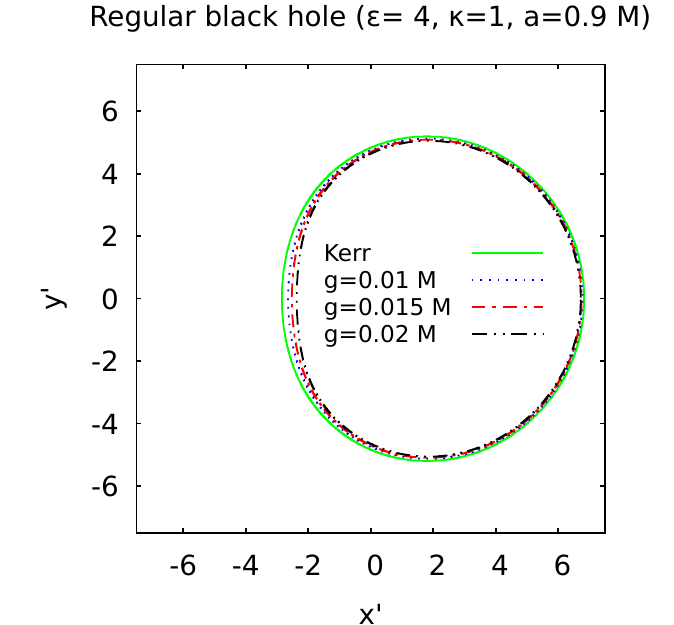}}
\caption{Top row: Shadows of a rotating regular black holes for different values of magnetic charge and rotation parameter, with $\kappa=1$ and $\epsilon=3$. Bottom row: Shadows of other rotating regular black holes for different values of magnetic charge and rotation parameter, with $\kappa=1$ and $\epsilon=4$. }
\label{Fig2}
\end{figure*}

\begin{figure}
  \centering
  \subfigure{\includegraphics[scale=0.5]{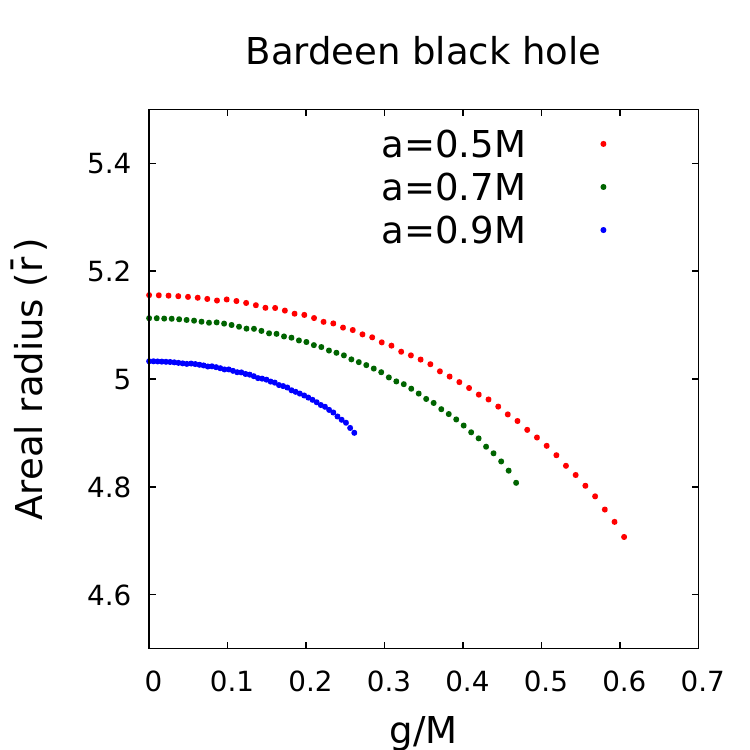}}
  \subfigure{\includegraphics[scale=0.5]{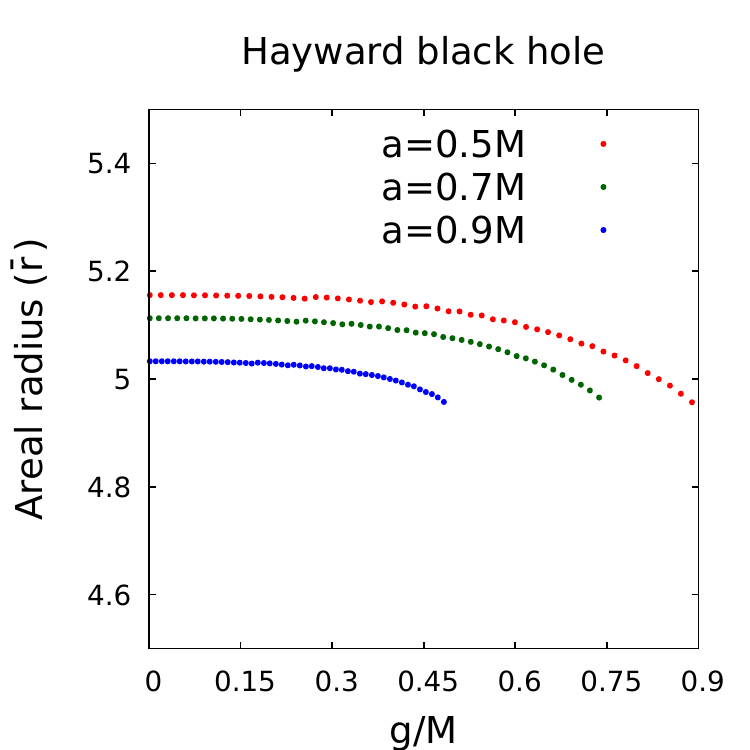}}
  \\
\subfigure{\includegraphics[scale=0.5]{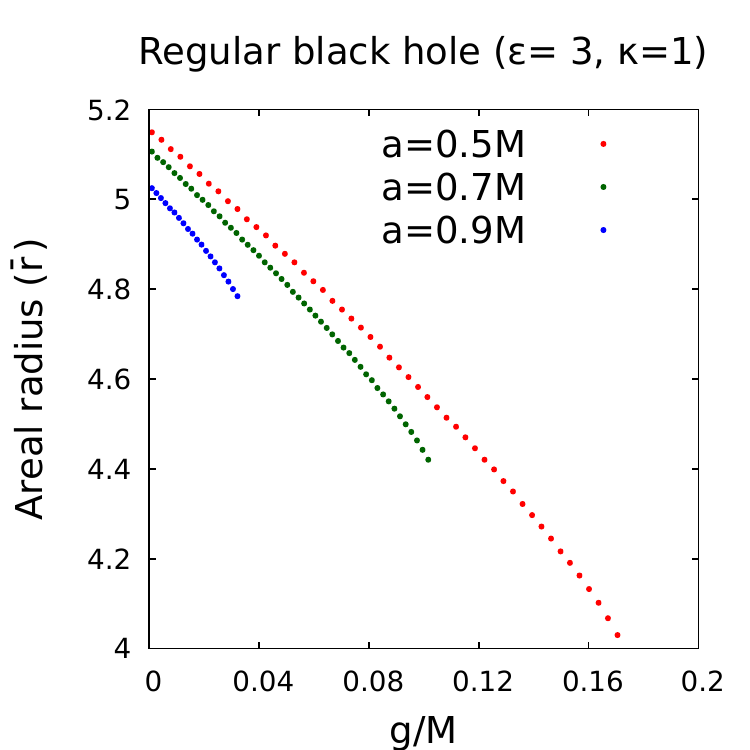}}
  \subfigure{\includegraphics[scale=0.5]{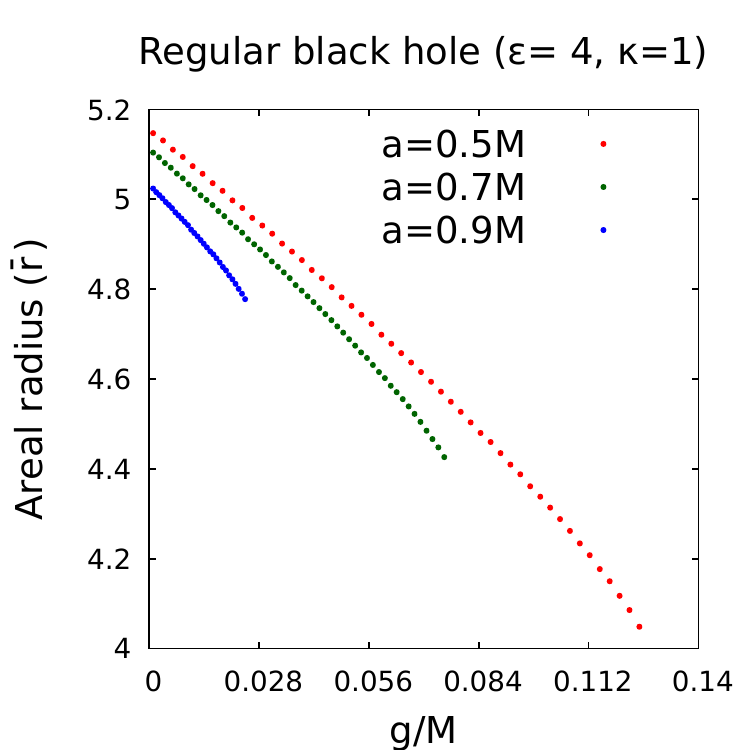}}
\caption{Areal radius $\bar{r}$ of Bardeen, Hayward and regular black holes with  $\kappa=1$ and $\epsilon=3,4$, as a function of $g/M$, for different values of rotation parameter $a$. We note that for the regular black holes presented here, the areal radius of the shadow decreases as we increase the parameter $g/M$.}
\label{Fig4}
\end{figure}

\begin{figure}
  \centering
  \subfigure{\includegraphics[scale=0.5]{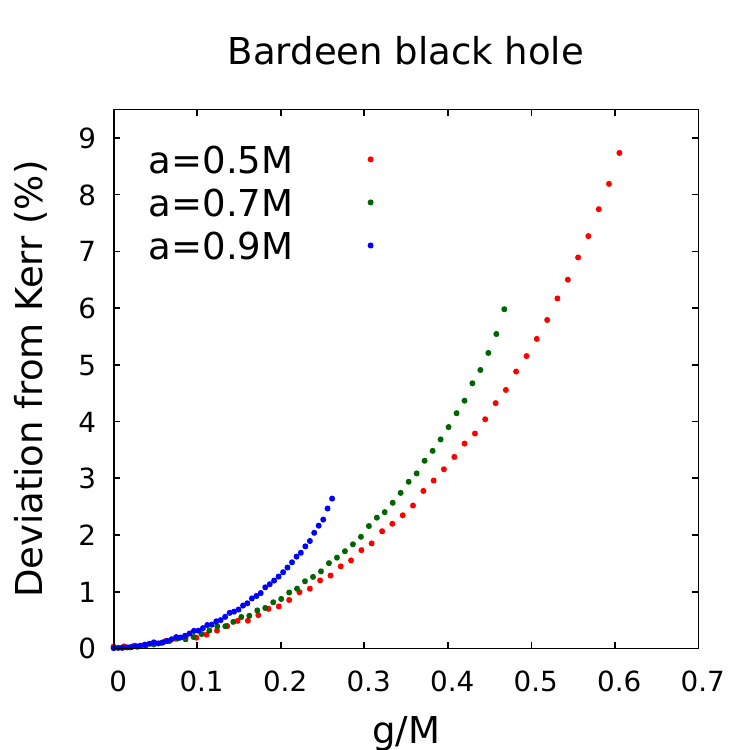}}
  \subfigure{\includegraphics[scale=0.5]{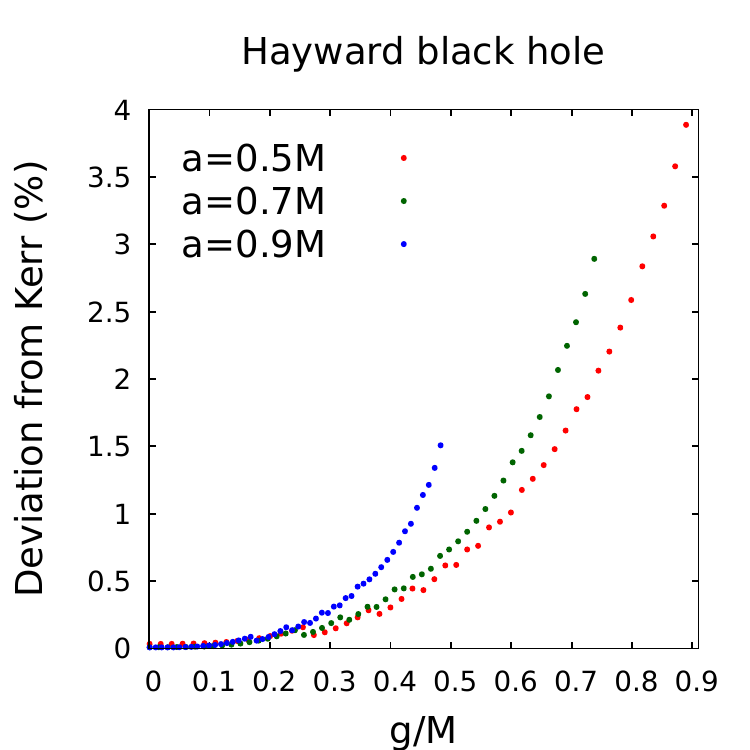}}
  \\
\subfigure{\includegraphics[scale=0.5]{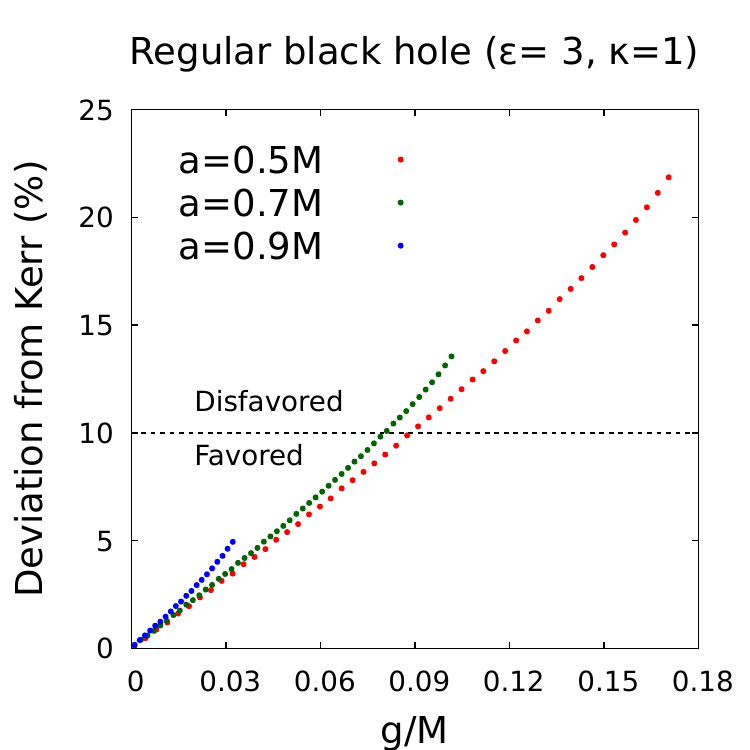}}
  \subfigure{\includegraphics[scale=0.5]{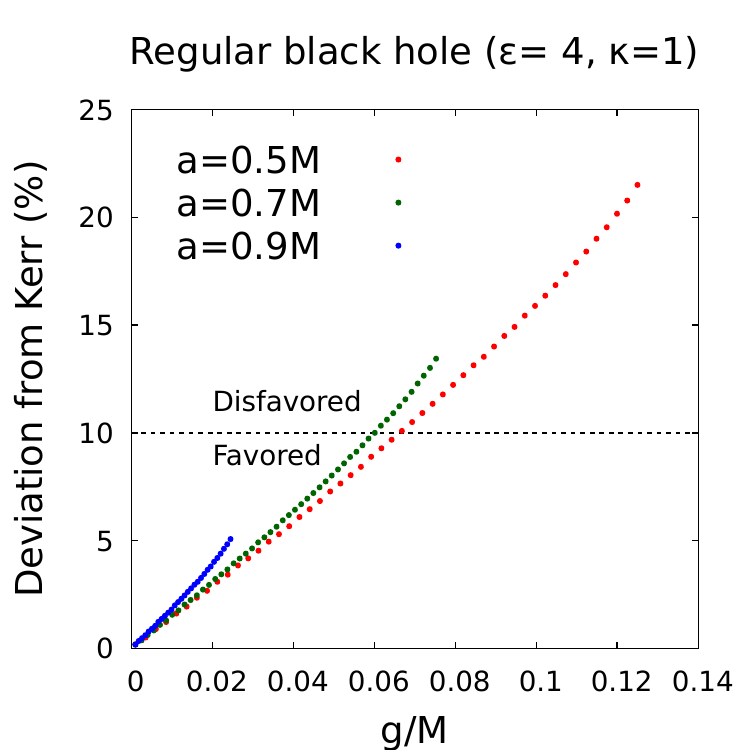}}
\caption{Relative deviation from the Kerr black hole as a function of $g/M$. We note that, for the regular black holes presented here, the relative deviation increases as we increase the parameter $g/M$.}
\label{Fig5}
\end{figure}

The shadow results are presented in Fig.~\ref{Fig2}, where we have chosen two types of regular black holes with $\kappa=1$ and $\epsilon=3,4$. From Fig.~\ref{Fig2}, we see that for fixed choices of $\kappa$, $\epsilon$ and $a$, the size of the shadow decreases as we increase the value of $g$. We can perform a quantitative analysis of this decreasing behavior by defining the areal radius $\bar{r}=\sqrt{\mathcal{A}/\pi}$, where $\mathcal{A}$ is the shadow area in the observer plane $x'-y'$. The areal radius $\bar{r}$ is well defined even for noncircular shadows. In Fig.~\ref{Fig4}, we present the areal radius for Bardeen, Hayward and the regular black holes with  $\kappa=1$ and $\epsilon=3,4$, as a function of $g/M$. We note that for the regular black holes investigated here, the areal radius decreases as we increase the parameter $g/M$. Moreover, in Fig.~\ref{Fig5} we show the relative deviation from the Kerr black hole, using the areal radius.

The recent observations of M87* by the EHT collaboration resolved an asymmetric bright emission ring with a diameter angular scale of $42\pm 3 \mu\text{as}$~\cite{EHT}. Although this observation is consistent with the shadow image of a Kerr black hole, there is still an uncertainty of $\sim 10\%$. Therefore, we can take a conservative bound of $10\%$ for the relative deviation of the shadow size from a comparable Kerr black hole as a simple assessment of which models could still be consistent with the EHT observations.
From Fig.~\ref{Fig5}, we note that Bardeen and Hayward black holes might still be consistent with the present observations of M87*, since the relative deviation from Kerr black hole is less than $10 \%$. On the other hand, some of the regular black hole configurations with $\kappa=1$ and $\epsilon=3,4$, are disfavored, depending on the values of $a$ and $g/M$, as can be seen in Fig.~\ref{Fig5}.

\begin{figure}
  \centering
  \subfigure{\includegraphics[scale=0.1]{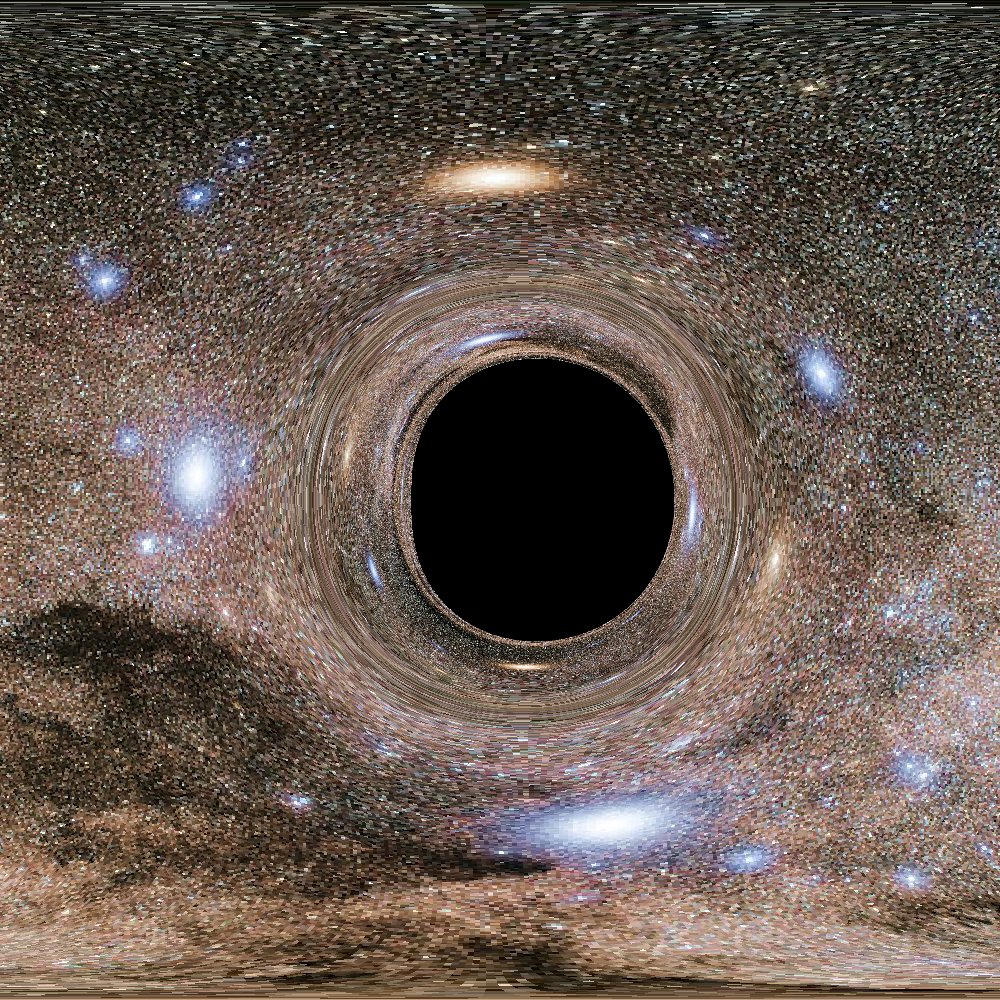}}
  \subfigure{\includegraphics[scale=0.1]{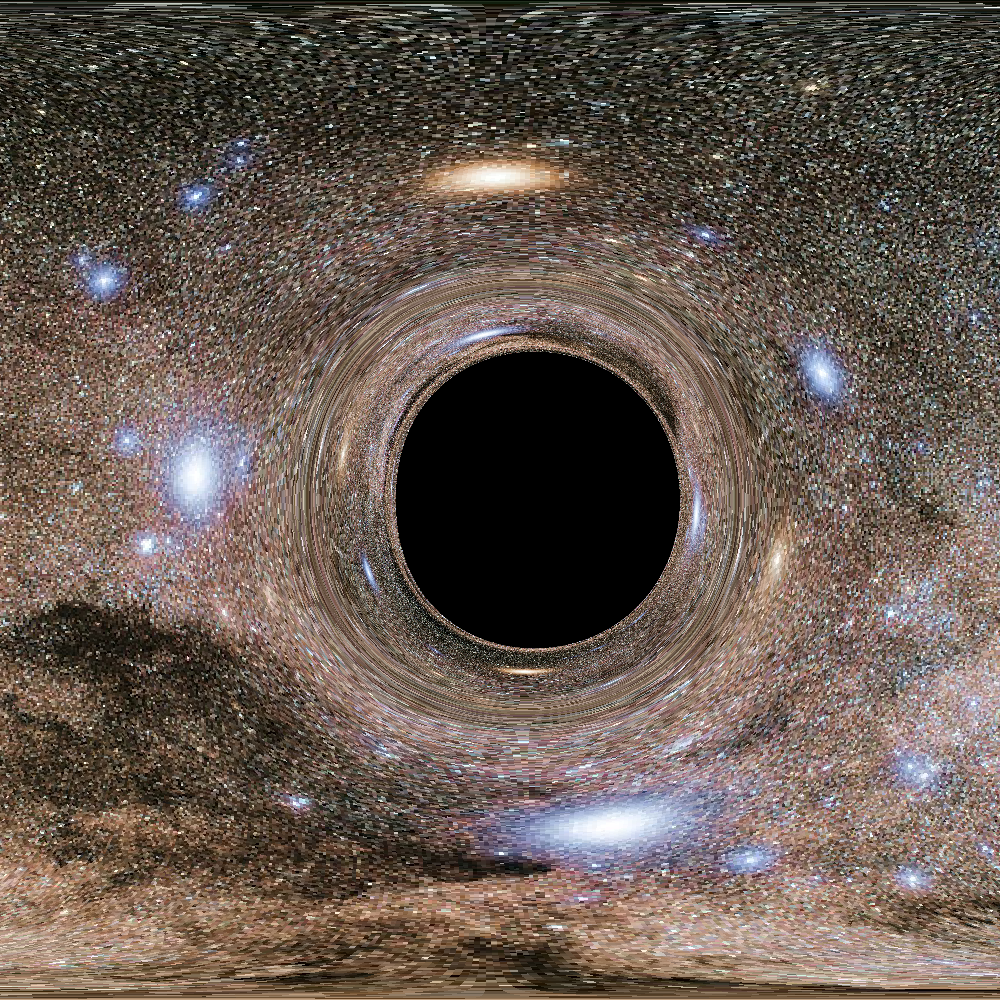}}
  \\
\subfigure{\includegraphics[scale=0.1]{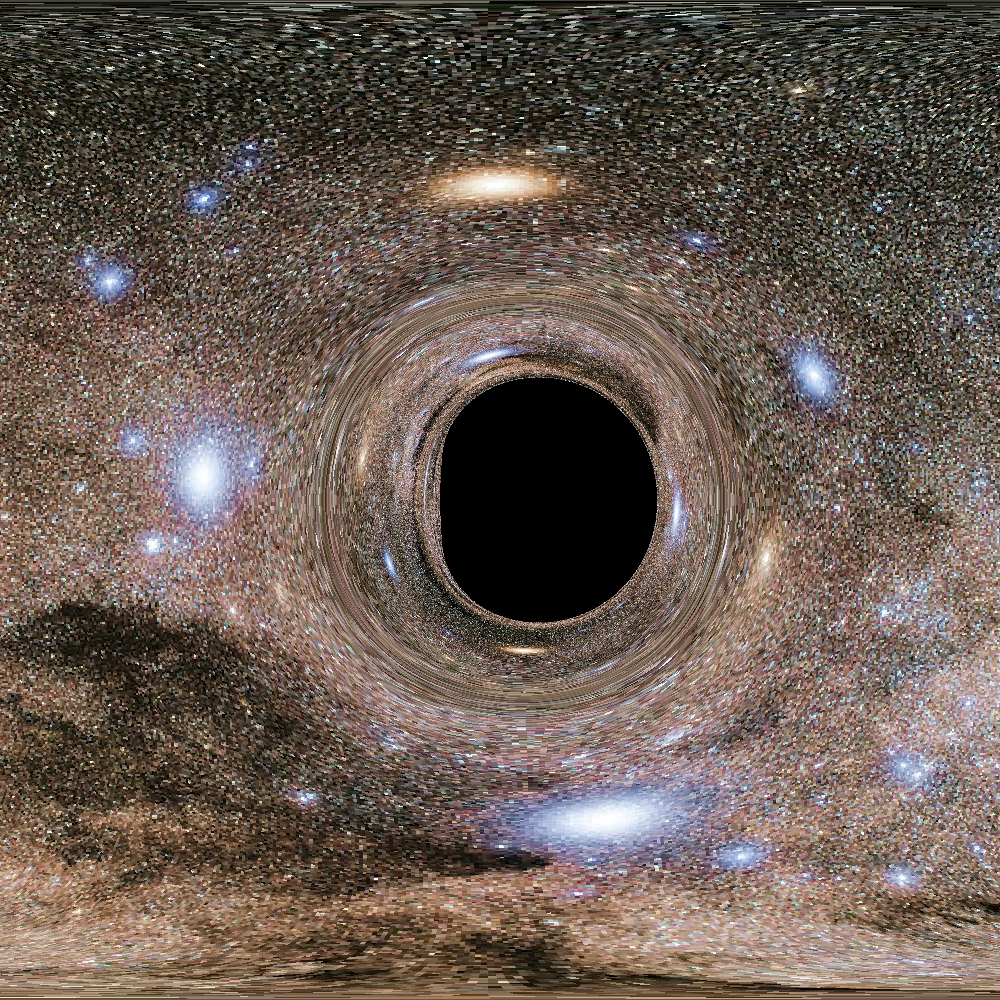}}
  \subfigure{\includegraphics[scale=0.1]{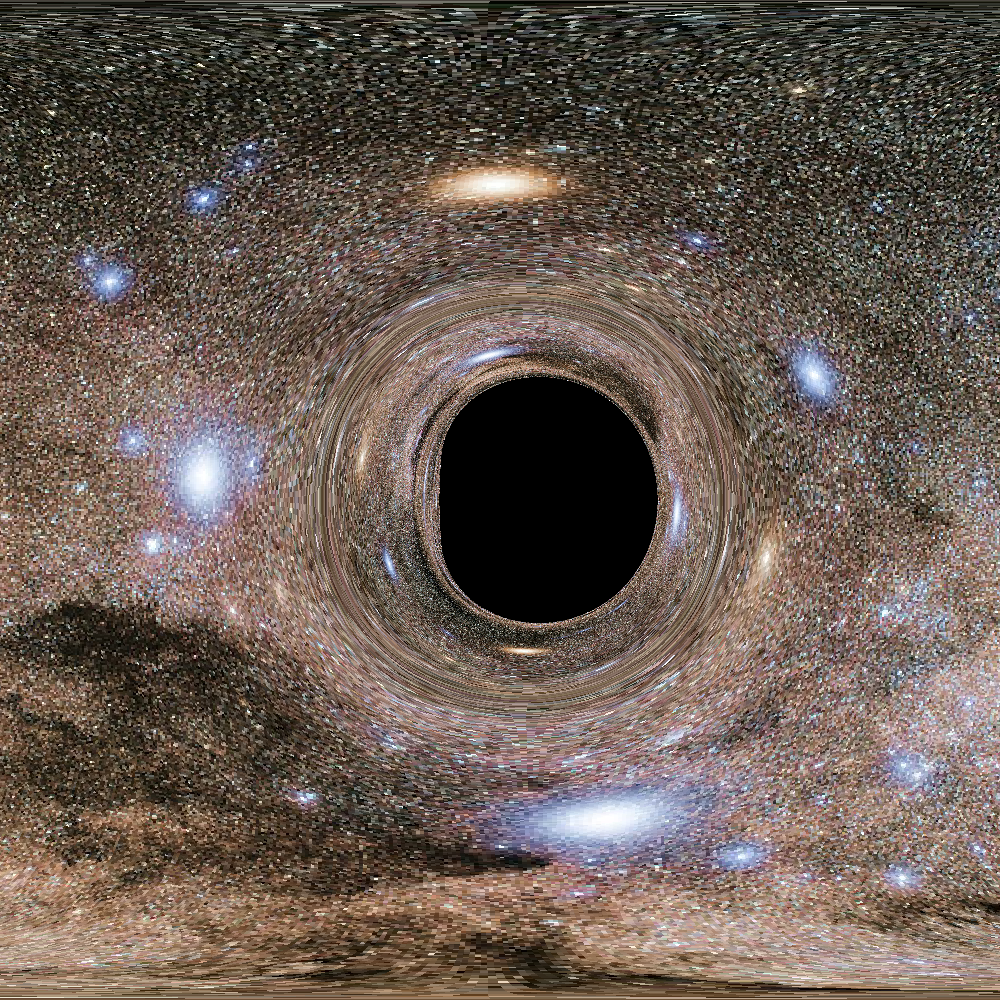}}
\caption{Shadows and lensing of regular black hole, under comparable observation conditions, obtained using backwards ray-tracing. Top row, left: Rotating Bardeen spacetime ($a= 0.5\,M$, $g= 0.617\,M$); Top row, right: Rotating Hayward spacetime ($a= 0.5\,M$, $g= 0.907\,M$). Bottom row, left: Rotating regular black hole with $\epsilon=3$ and $\kappa=1$ ($a= 0.5\,M$, $g= 0.174\,M$); Bottom row, right: Rotating regular black hole with $\epsilon=4$ and $\kappa=1$ ($a= 0.5\,M$, $g= 0.127\,M$). The background image can be found in~\cite{BKG_IMAGE}.
}
\label{Fig3}
\end{figure}

In Fig.~\ref{Fig3}, we present the shadows and gravitational lensing, obtained using backwards ray-tracing. We have chosen some of the rotating regular black holes solutions from Figs.~\ref{Fig1} and \ref{Fig2}. 
In order to produce Fig.~\ref{Fig3}, we evolved the light rays from the position of the observer backwards in time, until they reach the event horizon, or escape to infinity. We numerically integrated the null geodesic equation, with the initial conditions given by the photon's position and four-momentum with respect to a Zero Angular Momentum Observer (ZAMO), using a Dormand-Prince 5(4)  method~\cite{C++_book}.

\section{Conclusions}
\label{section:Conclusions}
The NJA has proved to be a useful solution generating technique in General Relativity, since it was introduced over half a century ago. In this paper we studied how a modified NJA~\cite{NJA_mod} impacts on the separability properties of the null geodesic flow in the resulting spacetime. 

Using the Hamilton-Jacobi formalism, we studied null geodesics of a generic spacetime generated with the MNJA, and found that the Hamilton-Jacobi equation \textit{always} admits separability. This is regardless of a certain ambiguity introduced by the MNJA, which turns out to be irrelevant for the study of the null geodesic flow. 

{We have also studied the propagation of light rays in the presence of a cold, pressureless and non-magnetized plasma. In the presence of a plasma, the light rays do not follow null geodesics. Still, the Hamilton-Jacobi equation in this case also admits separability, if the plasma frequency satisfies a given constrain, as given in Eq.~\eqref{Sep_cond_plasma}.}

We studied the spherical photon orbits in a generic spacetime generated by the MNJA, and found the condition for the existence of these orbits. Moreover, we analyzed the shadow cast by a generic spacetime generated with the MNJA and obtained the equations that describe the rim of the shadow cast by this rotating solution, as seen by an observer at infinity. We used our general result to obtain some of the particular cases previously presented in the literature. 

We studied the shadow of a generic magnetically charged regular black hole. The corresponding generic spacetime may describe the well known Bardeen and Hayward regular black hole solutions as particular cases.
In addition, it may also describe other regular black hole solutions. In one particular case, which we have taken as an example, the field approaches a Maxwellian field in the weak field limit. Although the shadows of Bardeen and Hayward black holes were already presented in the literature, we complemented the existing analyses by studying the gravitational lensing (using backwards ray-tracing methods) and the compatibility with the current observations of M87*. 
We concluded that for all the classes of rotating regular black hole we analyzed, the areal radius of the shadow is smaller
than the corresponding Kerr black hole (with the same values of the mass and rotating parameters). Moreover, for the Bardeen and Hayward regular black holes the relative deviation to Kerr are less than $10\%$, while for regular black holes with with $\kappa=1$ and $\epsilon=3,4$ this deviation can be greater than $10\%$, for some values of the rotation and $g$ parameters.

\begin{acknowledgements}
C. H. would like to thank E. Radu for many discussions. The authors thank Funda\c{c}\~ao Amaz\^onia de Amparo a Estudos e Pesquisas (FAPESPA),  Conselho Nacional de Desenvolvimento Cient\'ifico e Tecnol\'ogico (CNPq) and Coordena\c{c}\~ao de Aperfei\c{c}oamento de Pessoal de N\'{\i}vel Superior (Capes) - Finance Code 001, for partial financial support.
P.C. is supported by the Max Planck Gesellschaft through the Gravitation and Black Hole Theory Independent Research Group. This work is supported by the Center for Research and Development in Mathematics and Applications (CIDMA) through the Portuguese Foundation for Science and Technology (FCT - Funda\c{c}ao para a Ci\^encia e a Tecnologia), references UIDB/04106/2020 and UIDP/04106/2020. We acknowledge support  from the projects PTDC/FIS-OUT/28407/2017 and CERN/FIS-PAR/0027/2019.   This work has further been supported by  the  European  Union's  Horizon  2020  research  and  innovation  (RISE) programme H2020-MSCA-RISE-2017 Grant No.~FunFiCO-777740. The authors would like to acknowledge networking support by the COST Action CA16104. 
\end{acknowledgements}

\end{document}